\documentclass{article}
\usepackage{spconf,amsmath,graphicx}

\usepackage{booktabs}
\usepackage{tabularx}
\usepackage{hyperref}
\usepackage{multirow}
\usepackage{diagbox}

\usepackage{stfloats}

\newcommand\Mark[1]{\textsuperscript#1}

\makeatletter

\renewcommand\section{\@startsection 
{section}
{1}
{\z@}%
{-1.25ex \@plus -1ex \@minus -.2ex}%
{1.25ex \@plus.2ex}%
{\normalfont\Large\bfseries}
}

\renewcommand\subsection{\@startsection
{subsection}
{2}
{\z@}
{-1.5ex\@plus -1ex \@minus -.2ex}%
{1ex \@plus .2ex}%
{\normalfont\large\bfseries}}

\title{Beyond Neural-on-Neural Approaches to Speaker Gender Protection}

\name{Loes van Bemmel\Mark{1}\textsuperscript{,}\Mark{2}\textsuperscript{,}\Mark{3}, Zhuoran Liu\Mark{1}, Nik Vaessen\Mark{1}, Martha Larson\Mark{1}\textsuperscript{,}\Mark{3}}
\address{\Mark{1}Institute for Computing and Information Sciences, \Mark{2}Department of Artificial Intelligence,  \\\Mark{3}Center for Language Studies\\
Radboud University Nijmegen,
the Netherlands}

\begin{document}
\maketitle
\begin{abstract}
Recent research has proposed approaches that modify speech to defend against gender inference attacks.
The goal of these protection algorithms is to control the availability of information about a speaker’s gender, a privacy-sensitive attribute.
Currently, the common practice for developing and testing gender protection algorithms is “neural-on-neural”, i.e., perturbations are generated and tested with a neural network.
In this paper, we propose to go beyond this practice to strengthen the study of gender protection.
First, we demonstrate the importance of testing gender inference attacks that are based on speech features historically developed by speech scientists, alongside the conventionally used neural classifiers.
Next, we argue that researchers should use speech features to gain insight into how protective modifications change the speech signal.
Finally, we point out that gender-protection algorithms should be compared with novel “vocal adversaries”, human-executed voice adaptations, in order to improve interpretability and enable before-the-mic protection.
Code is available at \url{https://github.com/Loes5307/VocalAdversary2022}
\end{abstract}
\begin{keywords}
adversarial speech, gender inference, neural classifiers, interpretablity, attribute inference
\end{keywords}
\section{Introduction}
\label{sec:intro}

Recently, researchers have proposed approaches to protect spoken audio against gender inference attacks~\cite{gong2018, stoidis2022, Wu_etal_2021}.
The aim of these approaches is to protect speakers’ privacy by modifying their speech signal in a way that impedes the ability of a classifier to infer their gender.
The modified speech must retain its utility, which is generally taken to mean that it must remain understandable to people and have minimal impact on automatic speech recognition (ASR). 

This paper is motivated by our observation that current research devoted to developing and testing gender protection algorithms is overwhelmingly \emph{neural-on-neural}.
In other words, protective perturbations are generated with a neural approach and also tested against neural gender inference attacks. 
The main contribution of this paper are experimental results demonstrating the importance and usefulness of classic, non-neural speech features in attacking and analyzing gender protecting approaches, and motivating ``vocal adversaries", non-neural protection created by the human voice.

\begin{figure}
    \centering
    \includegraphics[width=0.9\columnwidth]{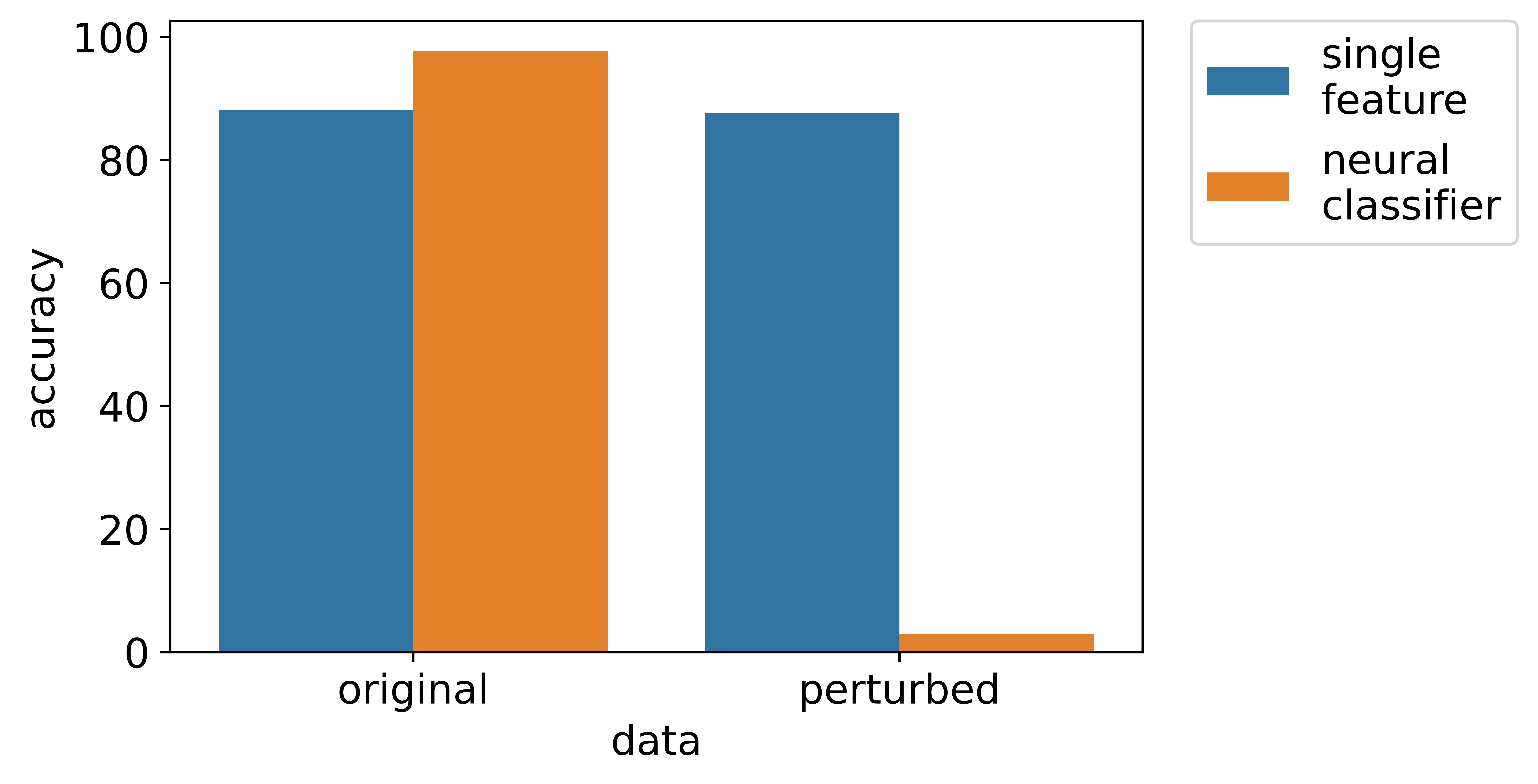} 
    \caption{Gender prediction accuracy on the VoxCeleb2 test set for 1) a single-feature classifier (linear ridge using mean pitch), and 2) a neural classifier (WavLM). Neural perturbations (reference model WavLM) do not affect the simple classifier, but highly impact the performance of the neural classifier (rightmost bar). Classifiers are trained on LibriSpeech.
   }
    \label{fig:barplot}
\vspace{-0.4cm}
\end{figure}

The importance of going beyond neural-on-neural approaches is illustrated by the gender inference attacks in Fig.~\ref{fig:barplot}. For the original, unperturbed data (left side), a simple classifier based on a single feature (mean pitch) is outperformed by a neural classifier (WavLM).
However, for data perturbed with a neural adversary (right side) the situation changes dramatically.  
The neural classifier can no longer correctly classify the speech samples, while the simple classifier maintains its performance. 
If researchers only test a neural classifier, as is common practice, e.g., ~\cite{gong2018, stoidis2022, Wu_etal_2021}, the vulnerability of the gender protection to a simple classifier based on speech features would go unnoticed.

The context of our work is the privacy threat represented by paralinguistic information inherent in speech, which has been gaining widespread attention recently~\cite{gong2018, tomashenko2020introducing}.
Gender is a privacy-sensitive attribute and concerns about gender classification being used for invasive or harmful targeting are growing~\cite{tuncay2015, turow2021}.
Further, targeting on the basis of binary notions of gender represents a grave reduction of the sociocultural concept of gender~\cite{lebourdais, zimman2021}.
We aim to move research on gender privacy protection closer to the real world.
This aim motivates our focus on protecting the raw signal, rather than speech representations, as has been studied by~\cite{Aloufi_etal2020, noe21_interspeech, stoidis2021}.
It also pushes us to look beyond neural attacks to extend the threat models under investigation to include non-neural modifications and before-the-mic scenarios. 
Initial efforts in these directions, and the closest work to our own,  is~\cite{Wu_etal_2021}, who tested a simple non-neural pitch shifting gender protection approach, and \cite{ahmed2022}, who investigate before-the-mic speech protection created by having speakers speak through a tube. 
In contrast, in our work we propose, for the first time, ``vocal adversaries'', speakers speaking with adaptations created using their own voices.

The paper is structured around three sets of experiments corresponding to the three ways in which we propose that it is important for speaker gender protection research to go beyond current neural-on-neural approaches.
In Sec.~\ref{sec:pointone}, we argue that researchers should be aware of the effect shown in Fig.~\ref{fig:barplot} and investigate gender inference attacks that use speech features historically developed by speech scientists, alongside  neural classifiers. 
In Sec.~\ref{sec:pointtwo}, we demonstrate how researchers can make use of speech features to seek insight into the ways in which protective modifications change the speech signal.
In Sec.~\ref{sec:pointthree}, we argue that neural approaches should not be considered the sole source of gender protection. 
Rather, we propose and introduce the promising ``vocal adversaries'', human executed voice adaptations.

\section{Beyond neural attacks}
\label{sec:pointone}
In this section, we demonstrate the vulnerability of typical neural gender protection to both neural classifiers as well as a classifier that uses classic speech features.

\subsection{Experimental Setup}

\begin{table}[]
\centering
\caption{Overview of datasets}
\vspace{0.05cm}
\label{tab:datasets-new}
\resizebox{\columnwidth}{!}{%
\begin{tabular}{@{}lrrrrrr@{}}
\toprule
 &  & \multicolumn{2}{l}{\#speakers} &  & \multicolumn{2}{l}{\#utterances} \\ \cmidrule(lr){3-4} \cmidrule(l){6-7} 
\textbf{data} & \textbf{avg. duration} & \textbf{F} & \textbf{M} & \textbf{} & \textbf{F} & \textbf{M} \\ \midrule
LS100h (training) & 12.7 s & 125 & 126 &  & 14 342 & 14 197 \\
LS960h (training) & 12.3 s & 1 128 & 1 1210 &  & 135 889 & 143 352 \\
vox (training) & 7.8 s & 2 312 & 3 682 &  & 397 032 & 694 977 \\ \midrule
Voxtest (testing) & 7.9 s & 39 & 79 &  & 10 711 & 25 526 \\ \bottomrule
\end{tabular}%
}
\vspace{-0.3cm}
\end{table}

\noindent \textbf{Data} Table~\ref{tab:datasets-new} presents the data for Male (M) and Female (F) speech. 
For the LibriSpeech~\cite{librispeech} data set, we use the train-clean-100 (LS100h) subset as well as all training data (LS960h).
LibriSpeech is known to be relatively `clean', meaning that the recordings do not contain significant background noise. 
For Voxceleb~\cite{voxceleb2}, which is known as a noisier dataset, the Voxceleb2 development set (designated as `vox') is used for training and the Voxceleb2 test set (`Voxtest') is used for testing. 
All audio recordings are in English, have a sample rate of 16kHz and are normalized to the [0,1] range. 
For testing we pad or cut each utterance to the first 6 seconds, similar to previous work~\cite{gong2018}.

\noindent \textbf{Neural classifiers}  We train two neural models for gender classification, which are later also used to create the neural perturbations.
M5~\cite{m5paper} is a convolutional neural network consisting of four convolutional layers with batch normalization, ReLU activation and a pooling layer, followed by a fully connected (FC) layer to classify gender.
WavLM~\cite{wavlm} is a self-supervised pre-trained transformer network that achieves state of the art results on the SUPERB benchmark.~\footnote{\url{https://superbbenchmark.org/leaderboard}} To perform gender classification with WavLM, we take the average of the output sequence from the transformer, and add three FC layers for classification.
M5 is trained from scratch on either LS100h, LS960h or vox for gender classification. 
WavLM's backbone is pre-trained on LS960h,\footnote{\url{https://huggingface.co/microsoft/wavlm-base}} then it is fine-tuned on either LS100h, LS-960h, or vox for the gender classification task.
We used Adam with a maximum learning rate (LR) of 1e-4 for M5 and 1e-5 for WavLM. 
All models are trained for 50k steps, with a cyclic learning rate schedule (triangular, 12.5\,k steps per cycle, minimum LR 1e-8). 
We use a batch size of 32 audio files and use raw waveforms as input. 
From each file, we randomly select a 3 second chunk for training. 
During evaluation, the first 6 seconds of the fragment is used.
For WavLM, we freeze everything but the last classification layers for the first cycle, before we unfreeze the transformer network. 
The feature extraction CNN of WavLM is kept frozen for the whole training duration. 

\noindent \textbf{Classic classifiers using speech features} 
Traditionally, speech research uses handcrafted features extracted from audio~\cite{bocklet2008}. 
These features have the advantage of having been designed by speech scientists, and, as such, reflect the underlying characteristics of speech. 
We extract 35 features with Praat~\cite{praat}, including number of Pulses, Periods and Voicebreaks. The degree of Voicebreaks, the fraction of Unvoiced parts, jitter (local, local absolute, rap, ppq5), shimmer (local, local dB, apq3, apq5, apq11), mean of the autocorrelation, Noise-to-Harmonics-Ratio (NHR), Harmonics-to-Noise-Ratio (HNR), mean and standard deviation of period and the min, max, mean, median and standard deviation of pitch.
We also included duration, intensity (min, max, mean, standard deviation), the fundamental frequency F0, first three formants and the centre of gravity.
These features were chosen for their widespread use in the acoustic community, as well as their interpretability with regards to speech production. 
We use the speech features in an SVM classifier trained on LS100h with a linear kernel to reduce overfitting.

\noindent \textbf{Feature selection} 
We select the top-10 features for the Female vs. Male classification using Recursive Feature Elimination (RFE) \cite{guyon2002} with a linear SVM. 
Any SVM used in feature selection is not used for final classification. 
SVM-RFE is a wrapper feature selection that utilizes the support vectors to discard the least important feature in each iteration, until a top-$n$ is left. The top-10 features can be seen in Table~\ref{tab:toptenfeats}. These features and the selection method will also be used in Sec~\ref{sec:pointtwo} for analysis of neural perturbations.

\begin{table}[]
\centering
\caption{The top-10 features ordered by importance for M vs. F classification obtained with SVM-RFE for LS100h.}
\vspace{0.05cm}
\label{tab:toptenfeats}
\resizebox{\columnwidth}{!}{%
\begin{tabular}{@{}lll@{}}
\toprule
\textbf{Feature}        & \textbf{Description}                & \textbf{Higher for:} \\ \midrule
pitch\_mean             & mean of pitch                       & Female                \\
autocor\_mean           & mean of autocorrelation             & Female                \\
nhr\_mean               & mean of Noise-to-Harmonics-Ratio    & Male                  \\
pitch\_std              & standard deviation of pitch         & Female                \\
pitch\_max              & max of pitch                        & Female                \\
intensity\_mean         & mean of intensity                   & Male                  \\
shimmer\_apq11          & shimmer computed with 11 neighbours & Male                  \\
shimmer\_apq3           & shimmer computed with 2 neighbours  & Male                  \\
intensity\_max          & max of intensity                    & Male                  \\
jitter\_local\_absolute & absolute jitter                     & Male                  \\ \bottomrule
\end{tabular}%
}
\vspace{-0.55cm}
\end{table}

\noindent \textbf{Neural gender protection}
We protect speakers' genders with neural perturbations created by applying Projected Gradient Descent (PGD)~\cite{madry2017} to trained neural networks (`reference models') in order to generate adversarial speech examples.
Compared to the single-step approach FGSM~\cite{goodfellow2015} used in~\cite{gong2018}, PGD is a stronger iterative version widely adopted in the adversarial machine learning community~\cite{tramer2020adaptive}.
PGD updates the perturbations iteratively:
\begin{gather}\label{eq:pgd}
     \boldsymbol{x}_{i + 1} = \boldsymbol{x}_{i} + \alpha \cdot \texttt{sign}(\nabla{J}(\boldsymbol{x}_{i}, y))
\end{gather}

where $\boldsymbol{x}_{i}$ is the perturbed waveform in iteration $i$, and $y$ is the label. 
$J$ denotes the Cross-Entropy loss.
Perturbations are generated by calculating $J$ on the reference model.
In each iteration, we clip values to $0.1$ to ensure that the perturbed speech is in a valid range to preserve utility on ASR systems. 
We use a perturbation rate $\alpha = 0.0005$, 100 iterations for perturbations generated with the M5 networks and 10 iterations for WavLM networks. The perturbations are generated using the first 6 second fragments of the audio.

Neural perturbations are considered to make a contribution to privacy if they reduce the accuracy of classifiers carrying out gender inference attacks either to random (0.5 in the case of balanced data), which protects a group of speakers, or to zero, which allows individual speakers to `flip' their gender.
The utility of the perturbed speech on ASR systems is measured by Word Error Rate (WER) of transcriptions.
Here, we compare the transcripts of Original speech to transcripts of protected speech using DeepSpeech 2~\cite{deepspeech2}.

\subsection{Results of Neural Attacks}
\begin{table}[]
\centering
\caption{Gender prediction accuracy on Voxtest protected with neural perturbations. Format: All (F/M). ``Ref model" specifies the reference model used to generate the perturbations. White-box accuracy is underlined. The Relative WER (rel WER) w.r.t. DeepSpeech 2 transcription of the `Original' data is reported as a measure of ASR utility.}
\label{tab:advspeechsota}
\resizebox{\columnwidth}{!}{
\begin{tabular}{@{}lcccc@{}}
\toprule
&\multicolumn{3}{c}{\textbf{Attack classifier}} &\\ 
\cmidrule{2-4}
\textbf{Ref Model} & \multicolumn{1}{c}{$\rightarrow$ \textbf{M5-LS100h}} & \multicolumn{1}{c}{$\rightarrow$ \textbf{M5-LS960h}} & \multicolumn{1}{c}{$\rightarrow$ \textbf{M5-vox}}&\textbf{rel WER}\\ \midrule
Original &  92.5 (95.5 / 91.3) & 95.5 (93.8 / 96.1) & 97.1 (94.6 / 98.1) & \\
\hline
M5-LS100h     & \underline{0.2} (\underline{0.3} / \underline{0.2})  & 11.7 (2.1 / 15.6) & 51.1 (37.7 / 56.6) & 28 \\
M5-LS960h   & 14.2 (4.1 / 18.3) & \underline{0.8} (\underline{0.1} / \underline{1.2}) & 31.9 (16.5 / 38.2)  &27\\
M5-vox     & 59.9 (49.5/ 64.1) & 48.8 (18.3 / 61.2) & \underline{0.3} (\underline{0.2} / \underline{0.3}) &27\\ \midrule
&\multicolumn{1}{c}{$\rightarrow$ \textbf{WavLM-LS100h}} & \multicolumn{1}{c}{$\rightarrow$ \textbf{WavLM-LS960h}} &  \multicolumn{1}{c}{$\rightarrow$\textbf{WavLM-vox}} &\\ \midrule
Original &  97.7 (97.5 /97.9) & 97 (97.6 / 96.8) & 99 (98.2 / 99.4)& \\
\hline
WavLM-LS100h&  \underline{3.0} (\underline{5.8} / \underline{1.8}) & 26.7 (26.6 / 26.7) & 35 (42.2 / 31.9)&32 \\
WavLM-LS960h&   47.1 (81.4 / 32.7) & \underline{3.9} (\underline{5.5} / \underline{3.3}) & 57.6 (76.6/49.7) &27\\
WavLM-vox&  85.1 (86.9 / 84.3) & 81.1 (73.8 / 84.1) & \underline{6.3} (\underline{9.4} / \underline{4.9}) &22\\ \bottomrule
\end{tabular}
}
\vspace{-0.3cm}
\end{table}

Table~\ref{tab:advspeechsota} shows the gender classification accuracy of the different neural models against conventional neural adversaries.
We use M5 and WavLM trained on different data sets as reference models to generate adversarial speech.
When the reference model has the same architecture and training data as the attack classifier (i.e., white-box adversaries), the classification accuracy on perturbed speech (on the diagonal and marked with underline) is low, as expected.
In other words, neural perturbations are effective against gender classification in a white-box setting. 
Interestingly, in quite a few cases with differences between the reference model and attack classifier (i.e., grey-box adversaries off the diagonal, which are more relevant for real-world settings) some protection is still possible.
Code and more results of different adversarial speech on different classifiers can be found in our GitHub repository.\footnote[3]{\label{github}\url{https://github.com/Loes5307/VocalAdversary2022}}

\subsection{Results of Speech Feature Attack}

\begin{table}[t]
\centering
\caption[Caption for LOF]{Gender prediction accuracy of an SVM using speech features on Voxtest protected with neural perturbations. Format: All (F/M). “Ref model” specifies the reference model used to generate the perturbations. Left column: SVM with top-10 features selected with SVM-RFE; Right column: SVM with all 35 features.}
\vspace{0.05cm}
\label{tab:loestableB2}
\resizebox{0.8\columnwidth}{!}{%
\begin{tabular}{@{}lcc@{}}
\toprule
&\multicolumn{2}{c}{\textbf{Attack classifier}} \\
\cmidrule{2-3}
\textbf{Ref Model}& \textbf{SVM top-10} & \textbf{SVM full} \\ \midrule
Original & 87.6 (91.3 / 86) & 79.7 (95.9 / 72.9) \\
\hline
M5-LS100h & 88.1 (90.7 / 87) & 80.9 (95.2 / 74.9) \\
M5-LS960h & 88 (90.7 / 86.9) & 82.7 (94 / 77.9) \\
M5-vox & 88.3 (90.9 / 87.2) & 82.9 (94 / 78.3) \\
WavLM-LS100h & 87.6 (91.1 / 86.1) & 82.7 (94.4 / 77.8) \\
WavLM-LS960h & 87.2 (91.2 / 85.5) & 81.8 (94.8 / 76.3) \\
WavLM-vox & 87.8 (91.3 / 86.3) & 81.7 (94.8 / 76.1) \\ \bottomrule
\end{tabular}%
}
\vspace{-0.3cm}
\end{table}

Table~\ref{tab:loestableB2} demonstrates SVMs with speech features (both top-10 and the full 35 features) are effective gender classifiers with an accuracy somewhat lower than neural models on the original data.
We also see that neural perturbations do not provide effective protection against speech-feature-based SVMs.
Neural-on-neural approaches would have missed this important vulnerability.
Interestingly, neural perturbations boost SVM accuracy slightly, suggesting enhancement of feature robustness. 
Additional experiments on LibriSpeech's test set indicate that this effect is data set specific, as can be seen on the GitHub page.\footnotemark[3]{}

\section{Analyzing neural perturbations}
\label{sec:pointtwo}

The high accuracies in Table~\ref{tab:loestableB2} suggest that the underlying aspects of speech that are important for speech-feature-based gender classifiers are missed by neural perturbations, or are not sufficiently modified.
In this section, we analyze the overlap between the features that are changed by neural-perturbation and the features that discriminate between genders.

\begin{table}[]
\centering
\caption{The intersection of the top-10 Male vs. Female features (listed in Table~\ref{tab:toptenfeats}) and Non-perturbed vs. Perturbed features (on Voxtest perturbed with different models, listed in `Ref model') }
\label{tab:loestableC}
\resizebox{\columnwidth}{!}{%
\begin{tabular}{@{}ll@{}}
\toprule
 \textbf{Ref model} & \textbf{intersection of top-10 features} \\ \midrule
\textbf{WavLM-LS100h} & autocor\_mean, nhr\_mean, intensity\_mean, shimmer\_apq3 \\
\textbf{WavLM-LS960h} & pitch\_mean, nhr\_mean, intensity\_mean, shimmer\_apq3 \\
\textbf{WavLM-vox} & pitch\_mean, nhr\_mean, intensity\_mean \\
\textbf{M5-LS100h} & shimmer\_apq3 \\
\textbf{M5-LS960h} & autocor\_mean, nhr\_mean, shimmer\_apq3 \\
\textbf{M5-vox} & pitch\_mean, autocor\_mean, nhr\_mean \\ \bottomrule

\end{tabular}%
}
\vspace{-0.3cm}
\end{table}

We use SVM-RFE to obtain the top-10 contributing features distinguishing perturbed vs. non-perturbed speech. Table~\ref{tab:loestableC} presents the intersection of these features and the top-10 features relevant for classifying gender given in Table~\ref{tab:toptenfeats}.
This list provides insight into which features are relevant for protecting against both neural and speech-feature-based attacks.

\section{Vocal adversaries}
\label{sec:pointthree}

\begin{table*}[b]
\vspace{-0.4cm}
\caption{Gender prediction accuracy of different models for four voice adaptations of vocal adversaries. Format: All (F/M). The WER per voice adaptation is reported on the left as `F/M' w.r.t. manual transcription.}
\vspace{0.05cm}
\label{tab:tabledversion5}
\resizebox{\textwidth}{!}{%
\begin{tabular}{lllllllll}
\hline
\textbf{WER} & \textbf{data} & \textbf{M5-LS100h} & \textbf{M5-LS960h} & \textbf{M5-vox} & \textbf{WavLM-LS100h} & \textbf{WavLM-960h} & \textbf{WavLM-vox} & \textbf{SVM-full} \\ \hline
35 / 31 & default & 100 (100 / 100) & 96 (92 / 100) & 63 (25 / 100) & 100 (100 / 100) & 100 (100 / 100) & 100 (100 / 100) & 100 (100 / 100) \\
67 / 89 & whisper & 50 (96 / 4) & 50 (0 / 100) & 50 (0 / 100) & 94 (100 / 88) & 94 (100 / 88) & 98 (96 / 100) & 50 (100 / 0) \\
25 / 38 & lowrobot & 100 (100 / 100) & 63 (25 / 100) & 50 (0 / 100) & 100 (100 / 100) & 100 (100 / 100) & 100 (100 / 100) & 100 (100 / 100) \\
27 / 42 & highrobot & 100 (100 / 100) & 98 (96 / 100) & 71 (42 / 100) & 100 (100 / 100) & 100 (100 / 100) & 100 (100 / 100) & 100 (100 / 100) \\
39 / 51 & overlyhappy & 50 (100 / 0) & 52 (88 / 17) & 63 (33 / 92) & 56 (100 / 13) & 71 (100 / 42) & 67 (100 / 33) & 50 (100 / 0) \\ \hline
\end{tabular}%
}
\end{table*}

Since intensity (a measure of loudness) and pitch were found to be important for protection of gender in speech in Sec.~\ref{sec:pointtwo}, our vocal adversaries leverage these features.
As a proof of concept, twenty voice adaptations inspired by the voice disguise literature \cite{zhang2008} were recorded by two speakers (one male, one female) reading the same passage of a story (``The Patchwork Girl of Oz" \cite{thepatchworkgirl}). 
Four of these adaptations are reported in Table~\ref{tab:tabledversion5}, where `default' refers to speaking normally, `lowrobot' and `highrobot' refer to speaking monotone like a robot in low and high pitch respectively and `overlyhappy' refers to speaking with a high pitch while smiling broadly. 
We found that some vocal adversaries, e.g., `overlyhappy', can impede a gender inference attack.
These results establish the viability of future interest on vocal adversaries.
Note that the WER for the vocal adversary is computed with DeepSpeech 2 against a manual transcription, and is relatively high for all speech reflecting out-of-vocabulary words in the story. 
For some adaptations, the WER does not drop, indicating that the vocal adversary does not necessarily have the privacy-utility trade off characteristic of neural perturbations. 
A full overview and more details of the vocal adversaries can be found in the GitHub repository.\footnotemark[3]{}

\section{Discussion and Outlook}
\label{sec:discussion}

In this paper, we have presented three sets of experiments whose results show three ways in which researchers should strive to move beyond neural-on-neural approaches when developing and testing speaker gender protection algorithms. 
In Sec.~\ref{sec:pointone}, we have shown that neural perturbations are often effective against neural gender inference.
This point is not particularly surprising, given the effectiveness of neural perturbations in the image~\cite{tramer2020adaptive} and the non-speech audio domain~\cite{esmaeilpour2019robust}.
In many cases, neural perturbations transfer from the reference model to a different attack classifier, which makes them seem highly effective at first consideration.
However, speech has an important distinction from images and non-speech audio: it is produced by the human body and enjoys regularities in underlying structure that reflect phonation and the resonances of the vocal tract.
The consequence is that a single feature can already be used to draw a useful decision boundary (Fig.~\ref{fig:barplot}), and an SVM based on classic speech features (Sec.~\ref{sec:pointone}) can perform nearly on par with a neural classifier.
Neural perturbations that protect gender apparently fail to influence the underlying characteristics of speech sufficiently, since they provide little protection against gender inference attacks based on speech features.
This conclusion is supported by the fact that SVMs are less successful in defeating neural perturbations in non-speech audio~\cite{esmaeilpour2019robust}. 

Moving forward, researchers in the area of attribute inference are well served to attempt to analyze the impact of perturbations on speech in more detail.
Some papers have moved in this direction by providing spectrograms of protected speech or by describing how it sounds~\cite{stoidis2022, Wu_etal_2021}.
Here, we suggest that speech features are useful for explaining the effect of perturbations in a way that is closely related to speech production. 
Future work should further evolve the analysis presented in Sec.~\ref{sec:pointtwo}.

Finally, we have shown in Sec.~\ref{sec:pointthree} that a vocal adversary has potential to defend against both neural and speech-feature-based attacks.
The human-executed voice adaptations that we tested required no special voice training, and open a new avenue for research on real-world protection against gender inference attacks in, i.e., smart speakers.

\section{Acknowledgements}
The first author was supported by a master-level ELLIS excellence fellowship at Radboud University Nijmegen. 

\bibliographystyle{IEEEtran}
\bibliography{references}
\end{document}